\documentclass[twocolumn, a4paper, prx, longbibliography]{revtex4-2}
\pdfoutput=1
\usepackage[utf8]{inputenc}
\usepackage[english]{babel}
\usepackage[T1]{fontenc}
\usepackage{amsmath, amsthm, amssymb}
\usepackage{enumitem}

\usepackage[usenames,dvipsnames]{xcolor}
\usepackage{xcolor}
\definecolor{myblue}{RGB}{29, 113, 184}
\definecolor{myred}{RGB}{217, 61, 61}
\usepackage{hyperref}
\hypersetup{
	colorlinks=true,
	linkcolor=myred,
	filecolor=magenta,      
	urlcolor=myblue,citecolor=myblue}
\usepackage[capitalize]{cleveref}
\usepackage{graphicx}

\usepackage[normalem]{ulem}

\begin{document}

\title{Learning to reset in target search problems}

\author{Gorka Mu\~noz-Gil}
\email{gorka.munoz-gil@uibk.ac.at}
\affiliation{Institute for Theoretical Physics, University of Innsbruck, Technikerstr. 21a, A-6020 Innsbruck, Austria}

\author{Hans J. Briegel}
\affiliation{Institute for Theoretical Physics, University of Innsbruck, Technikerstr. 21a, A-6020 Innsbruck, Austria}

\author{Michele Caraglio}
\affiliation{Institute for Theoretical Physics, University of Innsbruck, Technikerstr. 21a, A-6020 Innsbruck, Austria}

\begin{abstract}
Target search problems are central to a wide range of fields, from biological foraging to the optimization algorithms. Recently, the ability to reset the search has been shown to significantly improve the searcher's efficiency. However, the optimal resetting strategy depends on the specific properties of the search problem and can often be challenging to determine.
In this work, we propose a reinforcement learning (RL)-based framework to train agents capable of optimizing their search efficiency in environments by learning how to reset.
First, we validate the approach in a well-established benchmark: the Brownian search with resetting. 
There, RL agents consistently recover strategies closely resembling the sharp resetting distribution, known to be optimal in this scenario.
We then extend the framework by allowing agents to control not only when to reset, but also their spatial dynamics through turning actions.
In this more complex setting, the agents discover strategies that adapt both resetting and turning to the properties of the environment, outperforming the proposed benchmarks.
These results demonstrate how reinforcement learning can serve both as an optimization tool and a mechanism for uncovering new, interpretable strategies in stochastic search processes with resetting.
\end{abstract}
\maketitle

\section{Introduction}

Target search is a crucial task recurring in various scientific contexts, ranging from ecological foraging~\cite{sims2008scaling, raichlen2014evidence, brown2007levy} to optimizing search algorithms~\cite{yang2009cuckoo, molina2020comprehensive}. 
These problems often involve locating sparsely distributed targets in complex environments under various constraints.
In this context, resetting to a fixed initial condition has emerged as a powerful mechanism that allows to improve search efficiency in specific
circumstances~\cite{evans2020stochastic}. 
For instance, when performing optimization in complex loss landscapes, where one may get stuck in local minima, it may be advantageous to just restart the algorithm~\cite{luby1993optimal}. 
Another paradigmatic example is that of a Brownian walker searching for a target in free space, for which periodically restarting its position to the origin ensures a finite mean hitting time~\cite{evans2011diffusion}. 
In recent years, it has been shown that a particular resetting strategy, namely that in which a process is restarted at constant intervals, is typically optimal~\cite{pal2017first, chechkin2018random}.

However, while such sharp resetting is optimal in simplified scenarios, this is rarely the case in real-world applications, where search dynamics often involve additional complexities~\cite{de2020optimization, tal2024smart, plata2020asymmetric, kim2025emergence}. 
The optimal strategy in such cases may depend not only on the properties of the environment, such as the target's location and spatial constraints, but also on the capabilities of the agent performing the search~\cite{paramanick2024uncovering, blumer2024combining, sunil2024minimizing}.
Depending on the specific process under consideration, the agent may have access to a broader set of actions, defining its dynamics within the search process~\cite{kusmierz2014first}. 
The interplay between these dynamics and resetting can lead to more efficient target acquisition strategies~\cite{church2024accelerating}. 
However, identifying optimal strategies is highly non-trivial and often extremely challenging, as targets are typically sparse, making the optimization of search strategies computationally demanding.

In this work, we propose a machine-learning-assisted framework for identifying efficient target search strategies under restart.
In recent years, machine learning (ML) has demonstrated remarkable success in tackling complex physics problems~\cite{carleo2019machine, cichos2020machine, dawid2022modern}.
In particular, the optimization of particle navigation by means of ML has recently gained great attention~\cite{kolling2017reinforcement, nasiri2023optimal, muinos2021reinforcement}.
The strength of these techniques lies not only in their capacity to solve such tasks but also to uncover interpretable solutions that provide better insights into the underlying phenomena. 
In this context, Ref.~\cite{munoz2024optimal}, showed that reinforcement learning (RL) agents, provided with minimal prior information and trained solely to maximize the number of targets acquired, were able to develop efficient search strategies by uncovering the intrinsic scales of the environment.

Building on this foundation, we extend the approach by introducing agents with the ability to reset, the time at which this occurs being learned during their training (\cref{sec:agents}). 
In a benchmark scenario where the RL agents diffuse freely and can only choose when to reset, we show that they converge to the known optimal sharp resetting strategy (\cref{sec:results_resets}). 
We then address a more complex scenario in which agents adapt both their resetting distribution and their spatial dynamics.
By analyzing the learned behavior of the agent, we uncover new strategies that differ significantly from the optimal sharp strategy and, most importantly, outperform it (\cref{sec:results_turn_resets}).

\section{Methods}

\begin{figure}
   \includegraphics[width=\columnwidth]{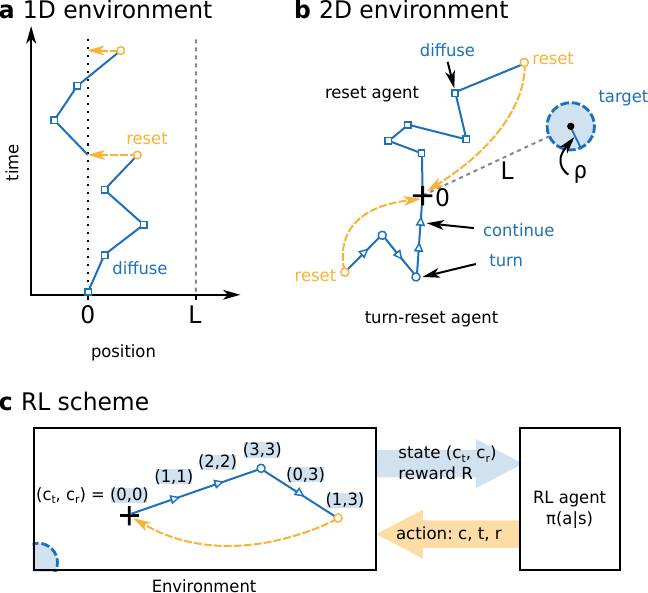}
   \caption{\textbf{RL formulation of the target search problem with resetting.}
    \textbf{a)} In 1D environments, the agent has two actions: diffuse (squares), where steps are sampled from a normal distribution with diffusion coefficient $D$, and reset (circles), which relocates the agent to the origin ($x=0$). A reward ($R=1$) is given upon reaching $x\geq L$.
    \textbf{b)} In 2D environments, we consider two agents: a \emph{reset} agent (as in 1D) and a \emph{turn-reset} agent, equipped with three actions: continue in the same direction, performing a step of constant length $d$ (triangles), turn by a random angle (blue circles) and reset (yellow circles). The target is positioned at distance $L$ from the origin and has radius $\rho$.
    \textbf{c)} The RL agent, exemplified here by a turn-reset agent, selects its next action $a$ —either continue (c), turn (t), or reset (r)— based on its policy $\pi$ and current the environment state $s$, defined in this case by two counters: steps since the last turn ($c_t$) and reset ($c_r$). See the main text for details.   
   }
    \label{fig:fig1}
\end{figure}

In this section, we present the main components of our study: the properties of the environment in which agents operate and the conditions under which targets are acquired (\cref{sec:envs}); the various agent types, distinguished by the set of actions they can perform (\cref{sec:agents}); and, finally, the reinforcement learning (RL) framework used to train agents in these environments (\cref{sec:rl}).

\subsection{Environment}
\label{sec:envs}
We consider an infinite environment, with a single target positioned at distance $L$ from the origin. Upon reaching the target, the agent resets and begins a new walk from the origin. 
We analyze two variants of this setup, one in one dimension (1D) (\cref{fig:fig1}a) and the other in two dimensions (2D) (\cref{fig:fig1}b). 
In 1D, the target is positioned at position $x = L$ and is considered acquired when the agent reaches a coordinate $x \geq L$.
In 2D, the target is still placed at a distance $L$ from the origin but the azimuthal angle is randomly selected in the interval $[0,2\pi)$.
In this second case, the target has a radius $\rho$ and is acquired whenever the agent enters the target area.
We consider that the target can be acquired even if the searcher trajectory crosses the target area in the middle of an integration step. 
As commonly considered, the target cannot be acquired while resetting to the origin.

To ensure non-trivial strategies, we consider scenarios where the target is positioned sufficiently far from the origin, discouraging strategies that forgo resetting entirely. 
This introduces a key challenge inherent to this class of search problems: as the target distance increases, successful acquisitions become rare events. 
Consequently, not only does the statistical characterization of the process become harder, but the effective training of learning agents is also increasingly difficult.

\subsection{Agents}
\label{sec:agents}
Given the previous environment, we consider two types of agents, distinguished by the set of actions they have access to (\cref{fig:fig1}). 

The first, called the \emph{Reset} agent in the following, is limited to two possible actions: performing a random motion (diffuse) or resetting to the origin (reset). More specifically, a diffuse action involves, in both one and two dimensions, sampling a step length from a Gaussian distribution and performing such step in a random direction, such that the agent would effectively perform Brownian motion with diffusion coefficient $D$.

The second agent, called the \emph{Turn-Reset} agent in the following, has access to three possible actions: reset, continue, and turn. 
The reset action again places the agent at the origin.
The continuing action allows the agent to perform a step in its current direction, with fixed step length $d$ (equal to $\rho$ throughout this work). 
The turning action enables the agent to change its direction to a new one defined by a random angle sampled from a uniform distribution.
The turning action also includes a step forward in the new direction to prevent the agent from becoming immobile during this process.

\subsection{RL algorithm}
\label{sec:rl}
To train an agent to maximize the number of targets acquired in the environments described above, we adapt the reinforcement learning (RL) procedure proposed in \cite{munoz2024optimal}. 
In this framework, the search problem is formulated as a Markov decision process (MDP), where the agent receives a state $s$ and selects an action $a$ based on the conditional probability distribution $\pi(a | s)$, typically referred to as the agent's \emph{policy} (\cref{fig:fig1}c). 

For the Reset agent, inspired by \cite{munoz2024optimal}, we define the state as an integer $s = c_r$, representing the number of steps taken since the last reset action. 
This \emph{counter} $c_r$ increments by one with each diffuse action and is set to zero whenever the agent performs the reset action.

In the Turn-Reset agent, the state is extended to include an additional counter, $c_t$, which tracks the number of steps taken since the last turn action. 
Consequently, the state is represented as $s = [c_t, c_r]$. 
In this case, the reset counter $c_r$ increments each time the agent performs either the continue or turn action and resets to zero after a reset action. 
Similarly, the turn counter $c_t$ increases by one whenever the agent performs the continue action but resets to zero when the agent executes either a turn or reset action.

The goal of the RL paradigm is to train a parameterized policy $\pi$ to maximize the expected rewards~\cite{sutton2018reinforcement}, which, in this context, corresponds to acquiring as many targets as possible during an episode lasting $T$ steps. 
This objective is equivalent to minimizing the time required to reach the target. 
A wide range of algorithms exists for training the policy $\pi$, with Q-learning, SARSA or REINFORCE being among the most widely used (see \cite{sutton2018reinforcement} for a comprehensive introduction). 
In this work, we employ Projective Simulation (PS)~\cite{briegel2012projective}, a related approach that offers specific advantages for the environment considered here. As noted previously, target acquisition is a rare event, making the rewards received by the agents extremely sparse. This means that a particular action may result in the target being reached only after many subsequent steps. 
For example, following a reset action, the agent must typically perform several continue and turn actions before eventually reaching the target.
Furthermore, the stochastic nature of the search process makes particular state-action pairs rather unrelated to the target-finding events.

The PS approach is particularly well suited for such scenarios~\cite{caraglio2024learning}, as it propagates the reward received at a given step back through all preceding state-action pairs.
In contrast, conventional RL algorithms typically propagate rewards only to the immediately preceding step, requiring significantly more iterations to effectively learn successful strategies in sparse environments like the ones considered here. 
The extent to which the obtained reward is propagated depends on the hyperparameters of the PS update rule, which needs to be tuned accordingly to effectively train the agents.
For further details on the methodology, we direct readers the accompanying Python library~\cite{rlopts}. The complete set of hyperparameters and training details used in this work can be found in the tutorial provided therein. For a full theoretical introduction to PS in the context of search strategies, we refer the reader to \cite{munoz2024optimal}.

Throughout this work, we consider completely unbiased initial policies, where all actions have equal probability at every counter value. Althoughugh it would be possible to encode prior knowledge into the initial policies—for instance, the simple notion that resetting at counters much smaller than the target distance $L$ is counterproductive—our goal is to evaluate whether the agent can learn this behavior autonomously, along with more complex strategies. Separately, we will also analyze the learned policies and, using both the insights gained from them and our own expert knowledge of the problem, refine the policies in a post-learning phase. We will refer to these refined strategies as \emph{expert-guided} policies.

\begin{figure*}
   \includegraphics[width=0.8\textwidth]{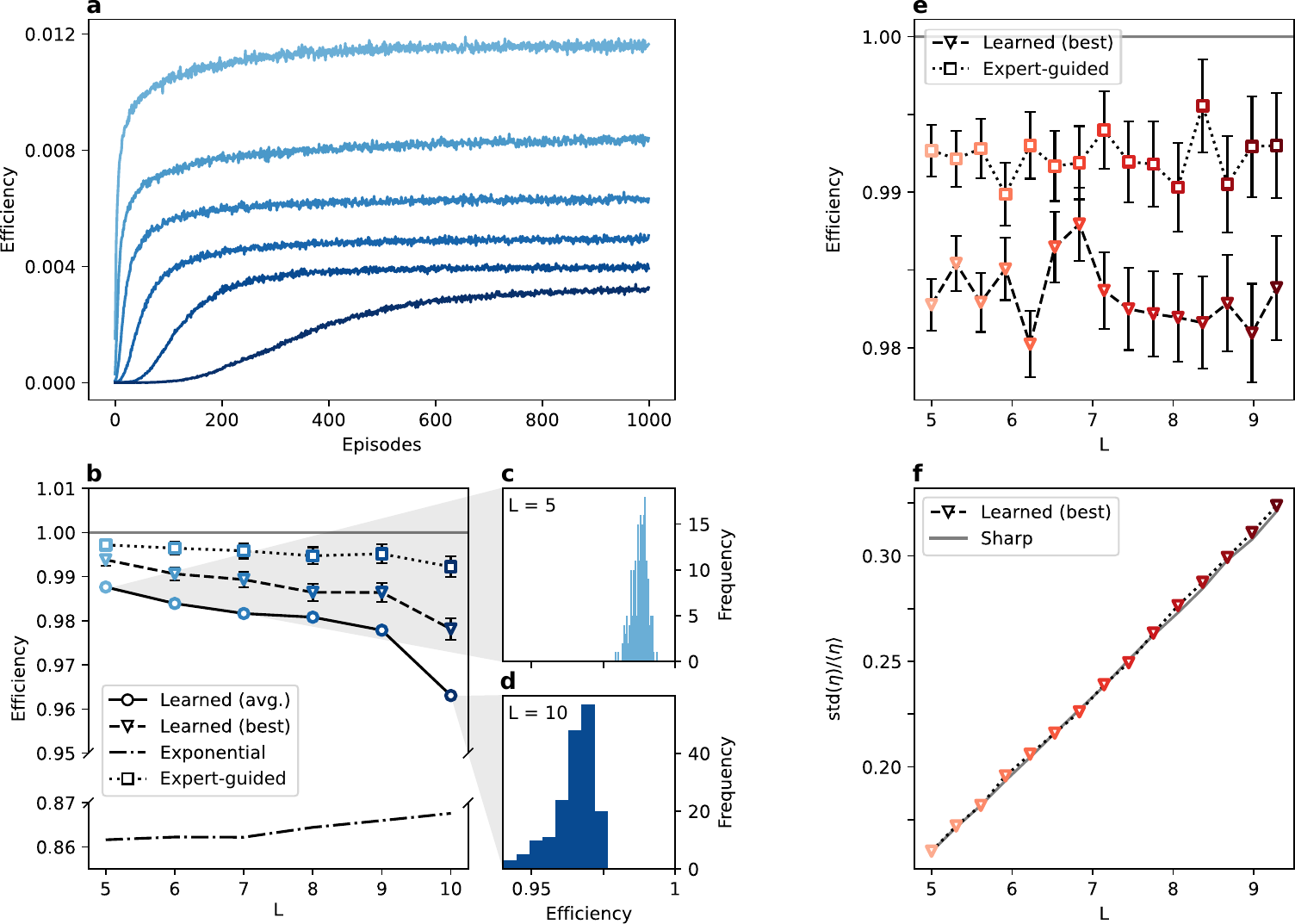}
   \caption{ 
   \textbf{Learned efficiency by the Reset agents} Overview of the efficiencies of the learning agents in 1D (panels a,b,c and d) and 2D (panels e and f) environments.
   \textbf{a)} Efficiency $\eta$ divided by the episode length $T$, as a function of the episode number, averaged over 190 agents. Each episode consists of $T = 5\cdot 10^3$ steps.
   \textbf{b)} Final efficiencies, divided by the sharp resetting efficiency $\eta_{\mathrm{sharp}}$, as a function of the target distance to the origin $L$. Each line corresponds to:  average over 190 agents (solid);  best agent (dashed);  exponential resetting rate (dot-dashed); best agent with a policy guided with expert knowledge (dotted).
   \textbf{c,d)} Distribution of efficiencies for the 190 learning agents for $L=5$ and $10$, divided by the sharp resetting efficiency $\eta_{\mathrm{sharp}}$.
   \textbf{e)} Final efficiencies of the learning agents in the 2D environment , divided by the sharp resetting efficiency $\eta_{\mathrm{sharp}}$,  as a function of the target distance $L$ divided by the target radius $\rho$. We consider here $\rho = 1$.
   \textbf{f)} Relative fluctuation of the efficiency at different $L$ for the sharp resetting strategy (solid) and the best learned strategy (triangles).
   }
    \label{fig:reset_eff}
\end{figure*}

\subsection{Baselines}
\label{sec:baselines}
To benchmark the efficiency $\eta$ of the learning agents, defined as the number of targets encountered per episode, we simulate well-known resetting strategies and compute their corresponding efficiencies. 
Across all experiments, we assume finite episodes of duration $T$. 
For sufficiently large $T$, the agent's behavior is expected to converge to an asymptotic value determined by the environment's properties.
However, since a finite $T$ imposes constraints on the environment (e.g., limiting the maximum excursion time to $T$), all agents --whether trained or set to baseline resetting distributions-- are evaluated under the same finite time conditions.

For agents equipped only with the reset action, it has been proved that the \emph{sharp} resetting strategy is optimal~\cite{pal2017first, chechkin2018random}. 
This strategy involves resetting at constant intervals of duration $\tau$, where the optimal value of $\tau^*$ depends on the environmental conditions, and particularly on the target distance. 
The optimality of the sharp resetting strategy provides a robust benchmark against which the performance of the learned strategies can be evaluated.
Our goal in this environment is therefore clear: to achieve an efficiency as close as possible to that of the sharp resetting distribution. 
Additionally, we compare the learned strategies to a commonly used probabilistic approach: resetting at each step with a constant rate, which corresponds to an \emph{exponential} resetting probability~\cite{evans2011diffusion}. While exact results for $\tau^*$ and the optimal resetting rate exists~\cite{chechkin2018random, evans2020stochastic}, due to the finite $T$ considered here, we select them via numerical optimization.

For agents capable of both turning and resetting, to the best of our knowledge, no strategy has yet been proven to be optimal. 
In this case, the strategy depends not only on the resetting distribution but also on the probability of the turning action, which is directly linked to the walker's step-length distribution. 
In the absence of resetting, this problem aligns with the well-known foraging problem, where the L\'evy walk hypothesis was initially proposed~\cite{viswanathan1999optimizing} but later widely challenged~\cite{levernier2020inverse,  benhamou2015ultimate}. 
The prevailing consensus now suggests that efficient strategies are those that leverage the intrinsic scales of the problem~\cite{ferreira2021landscape}. Ref~\cite{munoz2024optimal} showed that RL agents can indeed learn those scales and hence perform similar to the best known strategies in this scenario.

In our current environment, the problem has a single scale: the distance from the origin to the target minus the radius $\rho$. 
Drawing inspiration from the sharp resetting strategy discussed earlier, we extend this concept to create a baseline strategy for Turn-Reset agents. 
Specifically, we propose a strategy where the agent follows a deterministic turning rule, with a probability distribution $P(n) = \delta(n-N)$, indicating a turn precisely after $N$ steps. 
Similarly, as done above, the resetting distribution is also sharp, with the agent resetting always after $\tau$ steps.
We note that the most efficient strategies are always found at $\tau \geq N$. 
For each case studied, we determine the optimal values of $N$ and $\tau$ through numerical simulations. As we will show, both the $\tau$ and $N$ are found to be proportional to the single scale of the problem.

\section{Results}

In this section, we present the main results obtained for the two types of agents introduced earlier. 
We demonstrate that the RL method: (1) can achieve nearly-optimal resetting strategies and rediscover the sharp strategy without any prior information of its existence; and (2) Discovers an alternative strategy to the proposed sharp baseline, significantly outperforming it at specific target distances.
\begin{figure}
   \includegraphics[width=0.8\columnwidth]{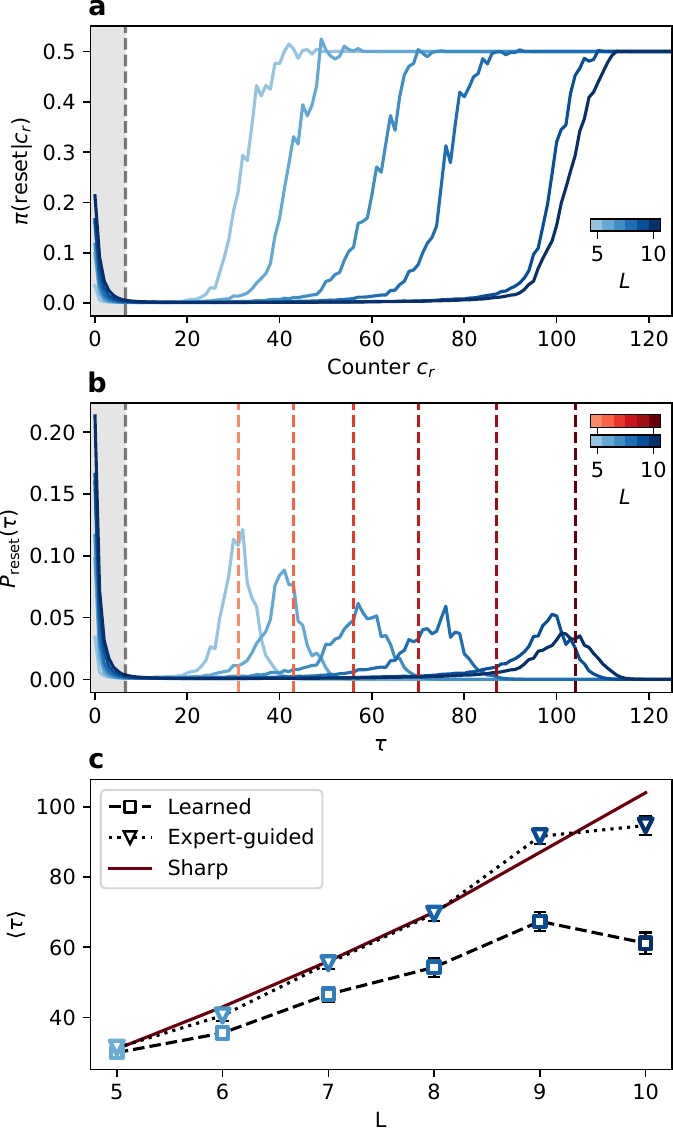}
   \caption{
   \textbf{Learned policies and resetting distributions of Reset agents in 1D environments} 
   \textbf{a)} Final learned policies, averaged over the 20 most efficient agents, for different target distances $L$ (blue shades). The gray area highlights the part of the policy set to zero to create the \emph{expert-guided} strategies.
   \textbf{b)} Resulting resetting distribution calculated from the previous policies. Vertical lines show the optimal reset of the sharp resetting strategy. The gray area shows the same as above.
   \textbf{c)} Mean resetting time for the learned and expert-guided strategies, calculated from the previous $P_{\mathrm{reset}}(\tau)$ distribution. Solid line shows the optimal reset for the sharp resetting strategy.
   }
    \label{fig:reset_policies}
\end{figure}

\subsection{Reset agents}
\label{sec:results_resets}
We begin by training agents that can choose between two actions: diffuse or reset. 
For both the 1D and 2D versions, we train agents at different target distances $L$ and analyze the resulting strategies. 
In \cref{fig:reset_eff}a, we present the evolution of the efficiency during training, averaged over 190 agents for various $L$ values in the 1D environment. 
As expected, smaller target distances $L$ yield higher efficiencies, as reaching the target is significantly easier. 
Conversely, learning efficient strategies for larger $L$ requires longer training times, as evidenced by the delayed plateau for $L = 10$.
This behavior arises because the probability of the agent reaching the target decreases substantially with increasing $L$, making it more challenging to learn the optimal sequence of actions.

Most importantly, as shown in \cref{fig:reset_eff}b, the agents successfully learn nearly optimal strategies (black solid line). 
Here, we plot the efficiency normalized by the optimal sharp resetting strategy. 
As shown, the efficiency remains above 96\% for all values of $L$, with the expected gradual decrease as $L$ increases, due to the increasing learning difficulty. 
Strikingly, the learned strategies significantly outperform the exponential resetting strategy (dash-dotted line), highlighting the ability of the agents to discover highly non-trivial solutions.
While the training process proves robust --ensuring that all agents achieve reasonably efficient strategies-- some variability is observed in their final efficiencies (\cref{fig:reset_eff}c and d). 
The dotted line shows the efficiency of agents whose policy has been refined following expert knowledge guidance (see \cref{sec:rl}). Further details on this will be given below.

When training agents in 2D environments, we observe the same result: agents successfully learn resetting strategies that are nearly optimal (\cref{fig:reset_eff}e). 
An important aspect to consider when evaluating resetting strategies is the standard deviation of their efficiency~\cite{pal2017first, eliazar2021tail}. 
From a practical standpoint, strategies with high mean efficiency but large fluctuations may be undesirable for certain applications, as they can lead to highly inefficient search runs. 
For the optimal sharp resetting strategy, it has been shown that the relative fluctuation $\mathrm{std}(\eta) / \eta$ satisfies $\mathrm{std}(\eta) / \eta \leq 1$~\cite{pal2017first}. 
In \cref{fig:reset_eff}f, we demonstrate that the relative fluctuation of the learned strategies closely follows that of the optimal sharp resetting strategy, further underscoring the robustness and reliability of the learned policies.

\subsubsection{Policies and learned resetting distributions}
Now that we have established that the agents' efficiencies are  comparable to that of the optimal sharp resetting, we examine the structure of the learned strategies. 
In~\cref{fig:reset_policies}, we show the agents' policies $\pi(\mathrm{reset} | c_r)$ averaged over the top-20 agents trained in the 1D scenario at different $L$ values (blue shades).
Before training, the policy is initialized at $\pi_0(\mathrm{reset} | c_r) = 1/2$, reflecting an equal probability of resetting or diffusing, and actually corresponding to an exponential strategy with constant (high) resetting rate.
For all $L$, a consistent trend emerges: through training, the probability of resetting decreases for smaller $c_r$ values. 
Beyond a certain point, $\pi(\mathrm{reset} | c_r)$ increases sharply and returns to the initial value of $1/2$. 
After this rapid increase, the policy stabilizes at the initial value because counters $c_r$ beyond this point are rarely encountered, as the agent would typically reset earlier. 
The location of this sharp transition depends on $L$, occurring at larger $c_r$ values for higher $L$. 
This behavior aligns with the requirements of the task: for larger $L$, the agent must allow more time between resets to successfully reach the target. 
Resetting too early (at $c_r \ll L$) would make it impossible to acquire the target.

A key strength of the framework presented here is its ability to transform the agent's learned policies into a probability distribution of resetting times, $P_{\mathrm{reset}}(\tau)$, as shown in ~\cref{fig:reset_policies}b. 
This transformation provides additional insights into the temporal resetting behavior of the agents. 
The resulting distributions reveal a non-zero resetting probability at short times $\tau$, which then decreases until a clear peak emerges at a specific $\tau$ value. 
This peak aligns closely with the optimal resetting time $\tau^*$ for the sharp resetting strategy (indicated by vertical dashed lines), known to scale as $(L-r)^2/D$ in the asymptotic limit $T\rightarrow\infty$~\cite{chechkin2018random}. 
This alignment demonstrates the agent's ability to approximate the optimal resetting times through its learned policy, even though the training does not explicitly enforce this alignment.

\begin{figure}
   \includegraphics[width=0.8\columnwidth]{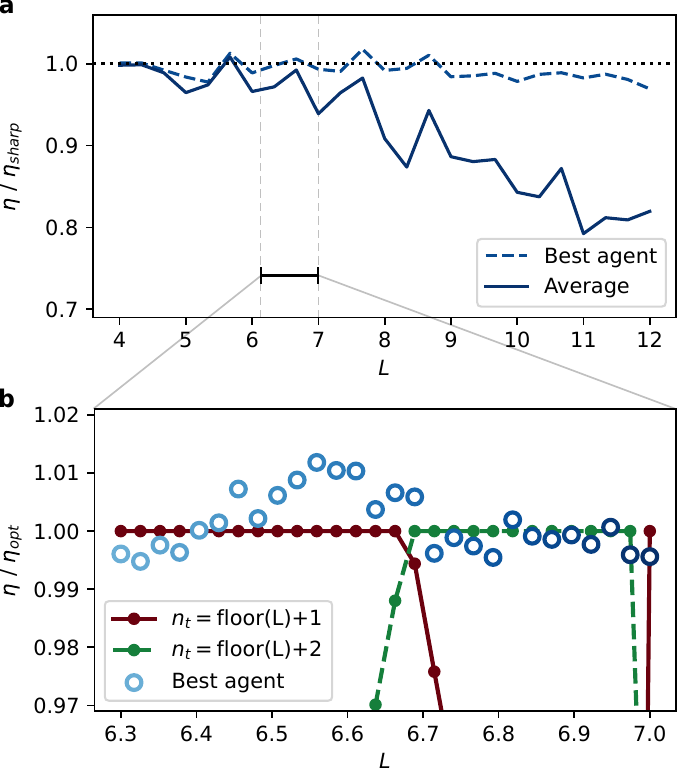}
   \caption{\textbf{Learned efficiency by Turn-reset agents} 
   \textbf{a)} Efficiency of the learning agents, normalized by the efficiency of the sharp strategy. Solid line: average over 80 agents. Dashed line: best agent at each $L$.
   \textbf{b)} Efficiency at finer resolution of $L$normalized by the best sharp strategy for each $L$: the sharp strategy with $n^* = \lfloor L \rfloor+1$ (solid red) for $L < 6.7$  and  $n^* = \lfloor L \rfloor + 2$ (dashed green) for larger $L$. Blue circles show the best agent out of 80 trained for each $L$. 
   }
    \label{fig:turn_reset_efficiencies}
\end{figure}

\begin{figure*}
   \includegraphics[width=\textwidth]{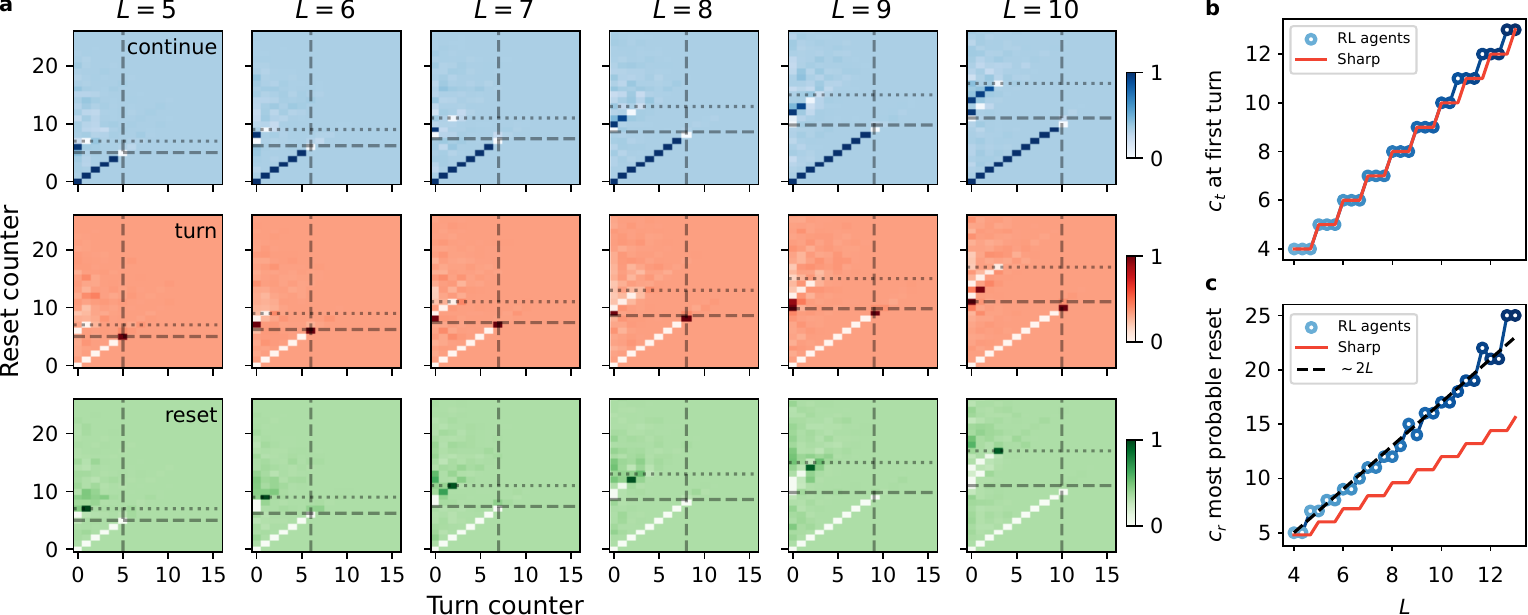}
   \caption{\textbf{Learned policies by turn-reset agents} 
   \textbf{a)} Every column shows the probability of performing the action continue (blue, upper panels), turn (red, center) and reset (green, lower) as a function of the current turn and reset counters ($c_t$ and $c_r$ respectively). We show here the policy of the best agent at each distance $L$ (see \cref{fig:turn_reset_efficiencies}). Dashed lines indicate the optimal turn and reset counters for the sharp baseline (horizontal and vertical, respectively). The horizontal dotted line marks the learned resetting time.
   \textbf{b)} Turn counter $c_t$ at which the initial turn is completed for the learning agents (blue) and the sharp strategy (red).
   \textbf{c)} Reset counter $c_r$ where the resetting action has the largest probability for the learning agents (blue). The red line shows the optimal $\tau^*$ for the sharp strategy and the black dashed line highlights a $2L$ scaling.
   }
    \label{fig:turn_reset_policies}
\end{figure*}

When examining both the policies and the resulting probability distributions, one notes that the agents wrongly assign a high resetting probability for small counters $c_r$. 
In the context of the current problem, resetting at $c_r \ll L$ is typically counterproductive, as the agent resets before traveling a distance $L$. 

This effect arises because the agents receive the reward much later in the state space (i.e., several steps are required before reaching the target, resulting in $c_r \gg 1$). Although the PS algorithm propagates the reward backward through all preceding actions that contributed to the current reward, it does so with an exponential decay based on the number of steps between the action and the reward. This makes learning the region $c_r \approx 0$ increasingly harder for larger $L$, as evidenced by higher $\pi(\mathrm{reset} | c_r = 0)$. Increasing the number of episodes steadily decreased this phenomena. However, to avoid the computational cost of the latter, and leveraging our expert knowledge about the problem, we refine the trained policies by setting $\pi(\mathrm{reset} | c_r < 5) = 0$ for all $L$ (gray are in \cref{fig:reset_policies}a and b). We will refer to this as the \emph{expert-guided} strategies.
This change led to an improvement in the resulting efficiency (\cref{fig:reset_eff}a, dotted line) and brought the mean resetting time of the expert-guided policies closer to that of the optimal sharp resetting strategy (\cref{fig:reset_policies}c, dotted line). 
This demonstrates the potential for combining learned strategies with domain knowledge to achieve further performance enhancements.

\subsection{Turn-Reset Agents}
\label{sec:results_turn_resets}

We now study the case in which agents are capable of learning when to turn, reset, or continue in a 2D environment. 
For the sharp baseline, the optimal strategy was found to be quite simple: perform a straight walk for $n^*$ steps of length $d$, turn, and then continue straight until resetting.
Interestingly, the optimal turning time $n^*$ is equal to $\lfloor L / \rho \rfloor +1$ whenever $L - \lfloor L \rfloor \lesssim 0.7 $ and $\lfloor L / \rho \rfloor + 2$ otherwise, remarking the geometrical subtleties of the problem. As $d=\rho = 1$, we will omit them in the following for simplicity. On the other hand, the optimal resetting time is found to be directly proportional to $L$, specifically $\tau^* \sim 1.2L$. Intuitively, this strategy enables the agent to reach the circumference from the agent where the target is located, followed by a short exploration of the surrounding area before resetting. Interestingly, this approach significantly outperforms a simpler strategy in which the agent resets immediately after traveling the distance $L+1$.

We then train agents in this scenario and plot their efficiency in \cref{fig:turn_reset_efficiencies}a. As shown, while the average efficiency across 40 trained agents decreases with increasing $L$ (solid line), there are consistently agents whose efficiency is comparable to that of the sharp strategy (dashed line). The former result is attributed to the fact that longer $L$ values require extended training times, which in our case is fixed to $5\cdot 10^4$ episodes for all $L$. Surprisingly, an oscillatory behavior emerges in both the average and best-agent efficiency. For the best agents, this means that at specific values of $L$—particularly those between half-integer and integer multiples—they significantly outperform the sharp strategy. To better understand this phenomenon, we train agents at a higher resolution of $L$, as shown in \cref{fig:turn_reset_efficiencies}b. The intersection of the solid red and dashed green lines marks the transition point where the optimal sharp strategy shifts from $n^* = \lfloor L \rfloor +1$ to $n^* = \lfloor L \rfloor + 2$. The agents consistently identify an efficient strategy across the entire range. Interestingly, they outperform the sharp strategy in the vicinity of its transition, suggesting the existence of a more efficient approach that cannot be replicated within the fixed constraints of sharp turns and resets.

To further investigate the source of this advantage, we examine the policies learned by the most efficient agents at various distances $L$ (\cref{fig:turn_reset_policies}a). 
The figure shows the probability of performing each action as a function of the reset and turn counters, $c_r$ and $c_t$ respectively. 
Since these probabilities are complementary, their values sum to one for any pair $(c_r, c_t)$.

As shown, the strategies learned at each $L$ exhibit a highly structured and consistent pattern. In all cases, the agent, positioned at the origin after resetting or encountering the target (i.e., $(c_r, c_t) = (0,0)$), begins with a very high probability of performing the continue action, hence increasing both counters by one. The continue action is then repeated until $(c_r, c_t) \approx (L, L)$ at which point the probability of performing the turn action becomes dominant (central panel), prompting the agent to turn (vertical dashed lines in \cref{fig:turn_reset_policies}a). This initial behavior mirrors that of the sharp strategy, where the agent executes a straight excursion of length $\sim L$ (\cref{fig:turn_reset_policies}b).
However, after this initial phase, the learned strategies diverge significantly from the sharp strategy. In all cases, the agents do not continue until resetting at the expected optimal sharp resetting time $\tau^*$ (horizontal dashed line in \cref{fig:turn_reset_policies}a). Instead, the agents perform few short steps, the number of which increases with $L$. After these, the agents eventually reset at approximately $c_t \sim 2L$ (horizontal dotted line), a notable deviation from the sharp resetting time of $\tau^* \sim 1.2L$ (\cref{fig:turn_reset_policies}c). This behavior shows that the agents exploit an alternative strategy that prioritizes additional exploration before resetting, which turns to be more effective at particular distances, as shown in \cref{fig:turn_reset_efficiencies}.

\section{Conclusions}
In this work, we demonstrated how machine learning (ML) can assist in discovering efficient strategies for target search problems with resetting—a fundamental problem across many fields~\cite{evans2020stochastic}. 
Resetting mechanisms are known to significantly enhance target acquisition, and we showed that reinforcement learning (RL) agents, equipped with actions controlling both their dynamics and resetting patterns, can achieve near-optimal strategies in known scenarios and aid researchers in discovering new strategies for unexplored problems.

In the paradigmatic Brownian search with resetting, the RL agents successfully converged to strategies closely resembling the \emph{sharp} resetting strategy, known to be optimal in such scenarios~\cite{pal2017first, chechkin2018random}. 
Additionally, we extended the framework to a more complex benchmark problem where agents control both step lengths and resetting distributions. 
By analyzing the learned policies, we devised novel strategies that outperformed the proposed sharp resetting a in specific geometric scenarios.

A key advantage of the proposed framework lies in its interpretability: the learned strategies are not only highly efficient but can also be extracted and understood within the established framework of resetting and target search problems. 
This interpretability bridges the gap between ML-driven optimization and theoretical insights, providing researchers with actionable strategies.

This study paves the way for applying the proposed methodology to more complex scenarios, particularly those where ``smart'' or adaptive strategies have proven successful~\cite{de2020optimization, tal2024smart, plata2020asymmetric, kim2025emergence, paramanick2024uncovering, blumer2024combining, sunil2024minimizing, kusmierz2014first, church2024accelerating}. 
By training RL agents in such scenarios, researchers can efficiently extract key features of the learned strategies and integrate them with their expert knowledge, enabling the discovery of new and robust solutions for a broad range of target search problems.

\bibliography{biblio}

\section*{Acknowledgments}
G.M-G. acknowledges support from the European Union.  M.C. is supported by FWF: P 35872-N.
This project was funded in part by the Austrian Science Fund (FWF) [WIT9503323]. This work was funded by the European Union (ERC, QuantAI, Project No. 101055129). Views and opinions expressed are however those of the author(s) only and do not necessarily reflect those of the European Union or the European Research Council. Neither the European Union nor the granting authority can be held responsible for them.

\end{document}